\documentclass[12pt]{article}
\usepackage{amssymb}
\begin{document}
\baselineskip=18pt
\pagenumbering{arabic}
\parskip1.5em
\thispagestyle{empty}
\vskip2.5em
\begin{center}
{\Large{\bf The coefficients of the Seiberg-Witten prepotential 
as intersection numbers (?)$^*$}}\\\vskip1.5em
R. Flume$^{1}$\hspace{3em}R. Poghossian$^{1\,\flat}$\hspace{3em}
H. Storch$^{1}$ \\ \vskip3em
$^{1}${\sl Physikalisches Institut der Universit\"at Bonn}\\
{\sl Nu{\ss}allee 12, D--53115 Bonn, Germany}\\
\end{center}
\vskip2em
\begin{abstract}
\noindent
The $n$-instanton contribution to the Seiberg-Witten prepotential 
of ${\bf N}=2$ supersymmetric $d=4$ Yang Mills theory is represented 
as the integral of  the exponential of an 
equivariantly exact form. Integrating out an overall scale and 
a $U(1)$ angle the integral is rewritten as $(4n-3)$ fold 
product of a closed two form. This two form is, formally, a 
representative of the Euler class of the Instanton moduli space 
viewed as a principal $U(1)$ bundle, because its pullback under 
bundel projection is the exterior derivative of an angular 
one-form. We comment on a recent speculation of Matone concerning 
an analogy linking the instanton problem and classical Liouville 
theory of punctured Riemann 
spheres.
\end{abstract} 
\vspace{2cm}
$^*$ {\small{To be published in the collection ``From 
Integrable Models to Gauge Theories'' (World Scientific, Singapore, 02 ) 
to honour Sergei Matinyan at the occasion of his 70'th birthday.}}\\
$^{\flat}$ {\small{on leave of absence from Yerevan Physics 
Institute, Armenia}}\\
{\small{
e-mail: flume@th.physik.uni-bonn.de\\
\hspace*{1.3cm}poghos@th.physik.uni-bonn.de\\
\hspace*{1.3cm}hst@mosaic-ag.com}}
\vfill\eject\setcounter{page}{1}

\section{Introduction}
One of the key events in the field of theoretical high energy 
physics in the last decade has been the proposal by Seiberg 
and Witten (SW) \cite{sw1}, \cite{sw2} of an exact formula for 
the so called prepotential of $d=4$, ${\bf N}=2$ supersymmetric 
Yang-Mills quantum field theories. The repercussions of the SW 
proposal were manifold: \\
(i) It is for the first time that exact statements on the dynamics 
of an interacting quantum field theory in four dimensions (beyond 
asymptotic freedom) are available; \\ 
(ii) The findings of SW have lead to new insights into the rich 
phase structure of supersymmetric Yang-Mills theories (\cite{s1}, 
\cite{abel} and references therein); \\
(iii) Soon after its discovery it became clear \cite{vafa}, 
\cite{lerche}, that the SW prepotential appears naturally in 
string theory as well as in field theory providing therewith a 
bridge between the two disciplines; \\
(iv) With the advent of Maldacena's proposal of the AdS/CFT 
correspondence \cite{maldacena} instanton calculations, both stringy 
\cite{green}, \cite{rossi} and field theoretical, \cite{dorey} 
became one of the mayor tools to check the proposal. 
The instanton calculations in the field theoretical context can 
be regarded as a continuation of the endeavor verifying the 
SW prepotential. \\ 
Such a central result as that of SW deserves for a rigorous proof. 
We hope to provide with the present article some modest steps into 
this direction. 

We will constrain ourselves to a discussion of the simplest 
possible setting for the appearance of an ${\bf N}=2$ prepotential. 
That is, we choose $SU(2)$ as underlying gauge group (spontaneously 
broken to $U(1)$) and restrict the field content to an ${\bf N}=2$ 
vector multiplet, denoted in the following by $\Psi$ \footnote{$\Psi$ 
is assumed to be composed of 
reduced $U(1)$-degrees of freedom due to symmetry breaking. $\Psi$ 
incorporates the breaking parameter $v$ : $\Psi=v+ \cdots$.}. 
The SW proposal  ${\cal F}(\Psi)$ as a function of $\Psi$ amounts under this 
circumstances to the ansatz for ${\cal F}$ as inverse of the  modular 
elliptic function $J$ \cite{erdelyi} with the following representation 
in terms of a power series expansion
\begin{equation}
{\cal F}(\Psi)=\frac{i}{2\pi} \log \Psi^2\log \frac{\Psi^2}{\Lambda^2} 
-\frac{i}{\pi}\sum_{n=1}^{\infty}{\cal F}_n \left(\frac{\Lambda}{\Psi}
\right)^{4n} \Psi^2, 
\label{Fexpansion}
\end{equation}   
where $\Lambda$ denotes a dynamical scale enforced by ultraviolet 
divergencies. The first (logarithmic) term on the r.h.s. of eq. 
(\ref{Fexpansion}) comprises the perturbative contributions which 
are the classical prepotential $\sim$ $\Psi^2$ and the one-loop 
contribution $\sim$  $\Psi^2 \log \Psi^2$. A non-renormalization 
theorem \cite{s2} implies that there are no other perturbative 
(higher loop) contributions to ${\cal F}(\Psi)$. 
The second term in the r.h.s. of eq. (\ref{Fexpansion}) with 
numerical coefficients ${\cal F}_n$ is of nonperturbative origin 
due to instanton configurations. The perturbative and non-perturbative 
pieces add up to the inverse elliptic modular function reflecting 
therewith a dynamically realized (S-) duality. For the arguments 
leading to the ansatz, (\ref{Fexpansion}), which might be called 
macroscopic - including in particular duality and a certain 
minimality assumption (which might be coined as the assumption of minimal 
analytical consistency) \cite{flume} - 
we refer to \cite{sw1} and \cite{sw2}. 

Our purpose in the present article is to reexamine the structure of 
the instanton calculations, that is, the ``microscopical'' approach 
towards a determination of the coefficients ${\cal F}_n$. 
We recall in Sec.~\ref{sec:II} part of the pioneering work by Dorey, 
Khoze and Mattis (DKM) \cite{dorey1}, \cite{dorey2}, \cite{dorey3}, 
leading to the determination of the one- and two-instanton 
coefficients \cite{dorey1} (for one instanton calculations see 
also \cite{finnel} - \cite{yung}) and also providing a suggestive ansatz
for the measure in the $n$-instanton moduli space\cite{dorey3}. It 
will be noted that the DKM measure has to be supplemented by the 
specification of a domain of integration, since the space of 
integration proposed by DKM containing redundant degrees of 
freedom is not orientable. In Sec.~\ref{sec:III} we show that ${\cal F}_n$ 
can be represented as integral 
of the top form of the exponential of an 
equivariantly exact form. Results equivalent to those in 
this section have been obtained in \cite{tanzini1} and \cite{tanzini2} within 
a technically different framework. Sec.~\ref{sec:IV} contains our main 
result, the representation of ${\cal F}_n$ as integral of a 
$(4n-3)$ fold product of a closed two-form. The two form is formally 
a representative of the Euler class associated with the ADHM moduli 
space viewed as a (principal) $U(1)$ bundle. 
In Sec.~\ref{sec:V} we relate our findings to an interesting 
speculation by Matone \cite{matone1} for a tentative analogy 
between classical Liouville theory on punctured Riemann spheres 
and the $d=4$ instanton problem. 

\section{The DKM results and the necessity of $SO(n)$ gauge 
fixing}\label{sec:II}
 
The coefficients ${\cal F}_n$ are given in terms of certain ``reduced'' 
matrix elements, that is, expressions in which the parameters 
corresponding to overall translation invariance are eliminated 
together with their supersymmetry partners. These matrix 
elements are represented by integrals over the supersymmetrized 
moduli space of $n$-instanton configurations - the Atiyah, Drinfeld, 
Manin (ADHM) \cite{adhm} parameters - 
\begin{equation}
{\cal F}_n=\int d\mu^{(n)}e^{-S^{(n)}}.
\label{modspaceint}
\end{equation}
$d\mu^{(n)}$ denotes here a particular measure on the ${\cal N}=2$ 
supersymmetrised ADHM moduli space and $S^{(n)}$ stands for an 
effective ${\cal N}=2$ action. We will quote the results of 
DKM concerning $d\mu^{(n)}$ and $S^{n}$ referring for detailed 
explanations to the articles \cite{dorey1}, \cite{dorey2} and  
\cite{dorey3}. 

The algebraic ADHM construction of selfdual $SU(2)$ Yang-Mills 
fields in four-dimensional Euclidean space - of integer valued 
topological charge $n$ - is based on an $(n+1)\times n$ matrix 
\begin{eqnarray}
a=\left( \begin{array}{ccc}
w_1 & \ldots & w_n \\
a'_{11} & \ldots & a'_{1n} \\
 & \ldots &  \\
a'_{n1} & \ldots & a'_{nn}
\end{array}\right) , \, \, \, a'_{ij}=a'_{ji}
\label{amatrix}
\end{eqnarray}
with quaternion valued entries $w_i$, and $a'_{1 \leq ij}; \,$ $1 \leq 
i,j \leq n$. 
We will enumerate the rows of the matrix $a$ by the integers 
$0, \ldots , n$, so that $a_{0i}\equiv w_i$ and $a_{ij}\equiv a'_{ij}$. 
$a$ is required to satisfy the quadratic constraint
\begin{eqnarray}
\Im m (\overline{a} a)_{ik}&\equiv& \frac{1}{2} \sum_{\lambda =0}^n
\left(\overline{a}_{\lambda i} a_{\lambda k} - 
\overline{a}_{\lambda k} a_{\lambda i} \right) \nonumber \\
&\equiv& \frac{1}{2} \left\{\overline{w}_i w_k  - \overline{w}_k w_i + 
\sum_{j=1}^n \left(\overline{a'}_{ji}a_{jk} - 
\overline{a'}_{jk} a_{ji} \right) \right\} =0, 
\label{constraints}
\end{eqnarray}
where $\overline{()}$ denotes quaternionic conjugation (for a matrix 
of quaternion this implies in addition also transposition)
\footnote{Our conventions concerning quaternions are the following: 
an arbitrary quaternion $x_{\alpha \dot{\beta}}$ can be 
decomposed in a basis of unit quaternions $x_{\alpha \dot{\beta}}= 
\sum_{\mu =1}^4 x_{\mu}\left(e^{\mu}\right)_{\alpha \dot{\beta}}$, 
$x_{\mu}$ real, $\left\{e^{\mu}\right\}=-i \sigma^{\mu}$, $\mu =
1,2,3$; $\sigma^4={\bf I}_2$; $\left(\bar{x}\right)^{\dot{\beta}\alpha}
=\epsilon^{\dot{\beta}\dot{\gamma}}\epsilon^{\alpha \delta}x_{\delta 
\dot{\gamma}}$; $\epsilon^{12}=\epsilon^{\dot{1}\dot{2}}=1$. 
The notions of imadginary ($\Im m$) and real ($\Re e$) parts, which 
we will use further on, should be understood with respect to this 
conjugation.}. 
Eq.'s (\ref{amatrix}) and (\ref{constraints}) together with the 
rank condition 
\begin{eqnarray}
\label{rankcondition}
\det \bar{\Delta} (x) \Delta (x) \ne 0, \\
\Delta (x) \equiv 
\left(
\begin{array}{c}
w_1, \cdots, w_n \\
a'_{ij}+x\delta_{ij}
\end{array}
\right) \nonumber 
\end{eqnarray}
for arbitrary quaternion $x$, 
provide the basis for an algebraic construction of (all) $SU(2)$ 
$n$-instanton configurations \cite{adhm}.

Chiral Weil zero-modes in the selfdual Yang-Mills background are 
constructed out of two $(n+1)\times n$ matrices ${\cal M}$, 
${\cal N}$
\begin{eqnarray}
{\cal M}=\left( \begin{array}{ccc}
\mu_1 & \ldots & \mu_n \\
{\cal M}'_{11} & \ldots & {\cal M}'_{1n} \\
 & \ldots &  \\
{\cal M}'_{n1} & \ldots & {\cal M}'_{nn}
\end{array}\right), \, \, \, \, \, 
{\cal N}=\left( \begin{array}{ccc}
\nu_1 & \ldots & \nu_n \\
{\cal N}'_{11} & \ldots & {\cal N}'_{1n} \\
 & \ldots &  \\
{\cal N}'_{n1} & \ldots & {\cal N}'_{nn}
\end{array}\right)
\label{mnmatrix}
\end{eqnarray}
with the (two component Weil-) spinor valued entries ($\mu_i= 
\left\{\mu_i^{\alpha}; \alpha=1,2\right\}$  and similarly for 
other entries) obeying the constraints
\begin{eqnarray}
\label{eq:fermconstraint1}
{\cal M}'_{ij}&=& {\cal M}'_{ji}, \, \, \, \, \, \,  
{\cal N}'_{ij} = {\cal N}'_{ji} \\
\overline{a}{\cal M}-\left(\overline{a}{\cal M}\right)^T&=&0, \, \, \, 
\overline{a}{\cal N}-\left(\overline{a}{\cal N}\right)^T=0.
\label{eq:fermconstraint2}
\end{eqnarray}
The undotted $2$-index of the quaternion $\overline{a}$ is 
paired here with the spinor label of ${\cal M}$ (${\cal N}$). 
For the subsequent discussion it is important to note that the 
data (\ref{amatrix})-(\ref{eq:fermconstraint2}) are redundant in 
the following sense: \\
Eq.'s (\ref{constraints}),(\ref{eq:fermconstraint2}) are invariant 
under $O(n)$ transformations 
\begin{eqnarray}
X \cdot a &=& \left(
\begin{array}{c} 
w \cdot X \\
X^{-1} \cdot a' \cdot X 
\end{array}
\right); \\
X \cdot {\cal M} &=& \left(
\begin{array}{c} 
\mu \cdot X \\
X^{-1} \cdot {\cal M}' \cdot X 
\end{array}
\right); \, \, \, 
X \cdot {\cal N} = \left(
\begin{array}{c} 
\nu \cdot X \\
X^{-1} \cdot {\cal N}' \cdot X 
\end{array}
\right), \\
X &\in& O(n) \,\,\,\, (X \cdot X^T=I_{n\times n}, \,\, X \,\, 
\mbox{real}) \nonumber
\end{eqnarray}
and $O(n)$ related data $a$ and $X \cdot a$ etc. give rise to 
the same vector potentials and Weil zero modes. 

The field content of $d=4$, ${\bf N}=2$ extended supersymmetric 
Yang-Mills field theory without matter fields can be read off 
from the Lagrangian
\begin{eqnarray}
{\cal L}_{{\bf N}=2}=tr \left\{-\frac{1}{4} F_{\mu \nu}F^{\mu \nu} 
+2{\cal D}_{\mu} \Phi^* {\cal D}^{\mu}\Phi 
\right.\qquad \qquad \qquad  \nonumber \\
+\left.\sum_{a}\left(i\bar{\lambda}_a \bar{\sigma}_{\mu} 
{\cal D}^{\mu} \lambda^a 
+g\Phi^* \left[\lambda_a ,\lambda^a \right]+g\Phi \left[\bar{\lambda}_a,
\bar{\lambda}^a\right]\right) +2g^2\left[\Phi^*, 
\Phi\right]^2\right\},
\label{ymlag}
\end{eqnarray}
which consists of a complex scalar field $\Phi$, two Weil 
spinors $\lambda_1$,$\lambda_2$ together with their chiral 
conjugates and a vector field $V_{\mu}$, all of which transform 
under the adjoint representation of the gauge group. 
The parts of (\ref{ymlag}) contributing to $S^{(n)}$ 
(eq. (\ref{modspaceint})) are the kinetic energies of the 
bosonic fields and the chiral Yukawa interaction: 
\begin{eqnarray}
S^{(n)}=\int d^4 x \,tr \left\{-\frac{1}{4} F_{\mu \nu}F^{\mu \nu}
+2{\cal D}_{\mu} \Phi^* {\cal D}^{\mu}\Phi +  
g\Phi^* \left[\lambda_a ,\lambda^a \right]\right\}.
\label{ymaction}
\end{eqnarray}  
The fields appearing on the r.h.s. are supposed to satisfy 
a reduced set of equations of motion
\begin{eqnarray}
\label{eq:eqmotion1}
F_{\mu \nu}&=&\tilde{F}_{\mu \nu}; \quad   
\mbox{($\tilde{F}$ denotes the dual of $F$)} \\
\label{eq:eqmotion2}
\bar{\sigma}_{\mu} 
{\cal D}^{\mu} \lambda^a &=&0;  \\
\label{eq:eqmotion3}
{\cal D}_{\mu}{\cal D}^{\mu} \Phi &=&- g \left[ \lambda_a ,\lambda^a \right]
\end{eqnarray}
and the symmetry breaking ($SU(2) \rightarrow SU(1)$) boundary 
condition  
\begin{equation}
\lim_{|x|\rightarrow \infty} \Phi (x)=v 
\label{vev}
\end{equation} 
with $v$ being a constant quaternion with vanishing real part but 
otherwise arbitrary. The terms, which have been omitted in the 
transition from eq. (\ref{ymlag}) to eq.'s (\ref{ymaction}) and  
(\ref{eq:eqmotion1}) -(\ref{eq:eqmotion3}), are of higher order 
in the coupling constant 
and are believed not to contribute to ${\cal F}_n$ \footnote{
A conceptually clear cut 
argument for this fact has not been given as far as 
we know.} The general solution of the equations 
(\ref{eq:eqmotion1}) -(\ref{eq:eqmotion3})
has been constructed by DKM in ref. \cite{dorey1}.  One has to 
insert this into $S^{(n)}$, eq. (\ref{ymaction}), and to write 
terms involving the scalar fields as a boundary contribution 
\begin{eqnarray}
\int d^4 x tr \left( {\cal D}_{\mu} \Phi^* 
{\cal D}^{\mu} \Phi -g \sqrt{2} \left[
\lambda , \lambda
\right] \Phi^{\dagger}
\right) \nonumber \\
=\int \partial_{\mu} tr \left(\Phi^* {\cal D}^{\mu} \Phi
\right) \nonumber \\ 
= \int dS^{\mu} tr \left(\Phi^* {\cal D}_{\mu} \Phi \right), 
\nonumber 
\end{eqnarray}
where for the first equality use has been made of the equation  of 
motion (\ref{eq:eqmotion3}). The integral in the last line is to be 
evaluated on the three sphere formally bounding Euclidean four-space. 
One obtains finally 
\begin{eqnarray}
\label{action}
\pi^{-2}S^{(n)}=\frac{8n}{g^2}+16|v|^2\sum_{k=1}^n |w_k|^2 
-8 tr_n \bar{\Lambda}{\bf L}^{-1}\left(\Lambda + \Lambda_{f} \right)  
+ 4\sqrt{2} \sum_{k=1}^n \bar{\mu}_k \bar{v} \nu_k ,
\end{eqnarray}
where $\Lambda$ and $\Lambda_f$ are $n\times n$ antisymmetric 
matrices 
\begin{eqnarray}  
\Lambda_{kl}&=&\bar{w}_k v w_l - \bar{w}_l v w_k ;\\
\Lambda_f&=&\frac{1}{2\sqrt{2}}\left(\bar{{\cal M}}{\cal N}- 
\bar{{\cal N}}{\cal M}\right),
\end{eqnarray} 
and ${\bf L}$ is a real operator acting on any skew symmetric 
$n\times n$ matrix $X$ by 
\begin{eqnarray} 
{\bf L}\cdot X = \frac{1}{2}\left\{X,W\right\} + 
\bar{a}'\left[a',X\right]- \left[\bar{a}',X\right]a', 
\label{loperator}
\end{eqnarray} 
where $W$ is the real valued symmetric $n\times n$ matrix
\begin{eqnarray}
W_{kl}=\bar{w}_k w_l +\bar{w}_l w_k.
\end{eqnarray} 
It can be shown that ${\bf L}$ is invertible for $a'$ and 
$w$'s satisfying the constraints (\ref{constraints}), 
(\ref{rankcondition}). 

The measure $d\mu^{(n)}$ is according to DKM \cite{dorey3} uniquely determined 
by the requirements of supersymmetry and the cluster property 
to be of the form  
\begin{eqnarray} 
d\mu^{(n)} =\frac{C_n}{Vol(O(n))} \prod_{i=1}^n d^4 w_i 
d^2 \mu_i d^2 \nu_i \prod_{i\le j}d^4 a'_{ij} 
d^2 {\cal M}'_{ij} d^2 {\cal N}'_{ij} 
\prod_{i<j}d\left({\cal A}_{tot}\right)_{ij} \nonumber \\ 
\times \prod_{i<j}
\delta \left(\left({\bf L} \cdot {\cal A}_{tot}
-\Lambda -\Lambda_f\right)_{ij}\right)
\delta^{(3)}\left(\frac{1}{2}\left(
\left(\bar{a}a\right)_{ij}-\left(\bar{a}a\right)_{ji}
\right)\right) \nonumber \\ 
\times \delta^{(2)}\left(\left(\bar{a}{\cal M}\right)_{ij}-
\left(\bar{a}{\cal M}\right)_{ji}\right) 
\delta^{(2)}\left(\left(\bar{a}{\cal N}\right)_{ij}-
\left(\bar{a}{\cal N}\right)_{ji}\right).
\label{measure}  
\end{eqnarray} 
The factor $1/Vol(O(n))$ takes care of the fact that the 
redundant $O(n)$ degrees of freedom have not been discarded 
from the integral. The normalization constant $C_n$ is fixed 
through the cluster condition and the normalization of the 
one-instanton coefficient. One easily recognizes in the 
last three groups of delta-functions on the r.h.s. 
of (\ref{measure}) the bosonic and fermionic constraints 
of eq.'s (\ref{constraints}), (\ref{eq:fermconstraint2}). 
The auxiliary variables ${\cal A}_{tot}$ can be integrated 
out leaving behind a factor $1/\det {\bf L}$. But the bouquet 
of variables (including ${\cal A}_{tot}$) chosen in (\ref{measure}) 
allows for a more direct verification of supersymmetry of the 
measure. (There should be, to start with, as many bosonic 
as fermionic differentials and $\delta$-functions for the sake 
of manifest supersymmetry.)

The DKM representation , eq. (\ref{modspaceint}), of ${\cal F}_n$ 
together with the specifications in eq.'s (\ref{action}), 
(\ref{measure}) has to be supplemented by an $O(n)$ gauge fixing  
condition since the larger 
space including redundant $O(n)$ degrees of freedom turns out to 
be non-orientable. To see the necessity of such a gauge fixing 
procedure we choose a gauge and verify a posteriori that the 
restriction is unavoidable to obtain a non-vanishing result 
for ${\cal F}_n$. We may consider for this purpose any representation 
built from the variables $w$ and $a'$ and impose on this gauge 
fixing conditions. To concretize the ideas let us consider the real 
symmetric matrix 
\[
Y_{ik}=\Re e \left(\bar{a}_{ij}a_{jk}\right)
\]
which transforms under the adjoint representation of $O(n)$. $Y$ 
can be brought through an $O(n)$ transformation into diagonal 
form with the diagonal elements arranged in increasing order of 
their absolute values \footnote{
This condition does not fix completely the gauge since some 
discrete transformations are still possible. Dropping
the factor $1/Vol(O(n))$ from (\ref{measure}) one has to take 
care of the latter leftover redundancies.}. There will not appear 
in this procedure a 
non-trivial Faddeev-Popov determinant because of supersymmetry. 
(The assertion will become evident from the deductions in the 
next section where we identify fermions with one-forms.)

The gauge fixing condition degenerates at places where two or more 
than two eigenvalues of $X$ coincide. There part of the $O(n)$ 
group is restored. In the generic case of two coinciding eigenvalues 
an $O(2)$ sub-group is revived. It means that we hit a Gribov 
horizon \cite{gribov}. Points at the two sides of the horizon 
of codimension one are 
related by a permutation (considered as an element of the group 
$O(n)$). This implies that the corresponding volume elements have 
opposite orientations what confirms the above claim 
\footnote{DKM avoid in \cite{dorey1} 
the horizon problem  by integrating the {\it modulus} of the integrand 
and the measure over the lager space.}. 

\section{Simplifications}\label{sec:III} 

To evaluate for general $n$ the integral (\ref{modspaceint}) with 
the measure (\ref{measure}) and the induced action, eq. 
(\ref{action}), appears to be a formidable task. The evaluation 
of ${\cal F}_2$, achieved by DKM, \cite{dorey1}, already seems, at 
least at first 
sight, to be miraculous. We propose in the following two 
subsections simplifications of the integrals, which we hope, will 
give some insight into the algebraic-geometric nature of the 
problem. The results of these two subsections have also 
been derived by Bellisai, Bruzzo, Fucito, Tanzini and Travaglini 
\cite{tanzini1}, \cite{tanzini2}, who use a technical approach different 
from our. 

\subsection{Fermions as differential forms}

It is a well known fact that fermion zero modes in selfdual Yang-Mills 
backgrounds are in correspondence with the fluctuation modes of the 
vector fields which makes it appearing natural to identify part or 
all of the cotangent space of the ADHM moduli with the Grassmann 
valued fermion zero modes. For 
the case of the ${\cal N}=2$ supersymmetry the correspondence of 
fermion modes and cotangent space turns out to be one-to-one. 

To start with we combine the Grassmannian spinor valued matrices 
${\cal M}$, ${\cal N}$ into a quaternion valued matrix denoted by 
${\cal P}$. Let $A^{\dot{\alpha}}$, $B^{\dot{\alpha}}$ be some 
two component ($\dot{\alpha}=1,2$) c-number spinors satisfying 
\begin{eqnarray}
A\cdot B\equiv A^{\dot{\alpha}} B_{\dot{\alpha}} \equiv 
A^{\dot{\alpha}} B^{\dot{\beta}} \epsilon_{\dot{\alpha}\dot{\beta}} 
=1 \nonumber \\
B_{\dot{\alpha}}=\epsilon_{\dot{\alpha}\dot{\beta}}
B^{\dot{\beta}}, \epsilon_{\dot{1}\dot{2}}=-\epsilon_{\dot{2}\dot{1}} 
=1, \nonumber
\end{eqnarray}
and define 
\begin{equation}
\left({\cal P}\right)_{\alpha \dot{\alpha}}={\cal M}_{\alpha} 
B_{\dot{\alpha}}-{\cal N}_{\alpha}A_{\dot{\alpha}}
\end{equation}
with the inverse relation
\begin{equation}
{\cal M}_{\alpha}=\left({\cal P}\right)_{\alpha \dot{\alpha}} 
A^{\dot{\alpha}}; \, \, \, \,
{\cal N}_{\alpha}=\left({\cal P}\right)_{\alpha \dot{\alpha}} 
B^{\dot{\alpha}}.
\end{equation}   
ADHM matrix labels are here suppressed. The constraints 
(\ref{eq:fermconstraint2}) read in terms of the new variables as 
\begin{equation}
\bar{a}{\cal P}-\left(\bar{a}{\cal P}\right)^T=0,
\label{newfermconstraint}
\end{equation}
and the fermionic $\delta$-functions appearing in the measure 
(\ref{measure}) read as 
\begin{equation}
\prod_{i<j}\delta^{(4)}\left(\left(\bar{a}{\cal P}\right)_{ij}-
\left(\bar{a}{\cal P}\right)_{ji}\right). 
\label{fermdelta}
\end{equation}
It is easily seen that the imaginary parts of the 
fermionic constraints (\ref{newfermconstraint})
are automatically fulfilled if one substitutes for ${\cal P}$ 
any $O(n)$-covariant derivative of $a$, 
\begin{eqnarray}
{\cal P}=\left(d+X\right)a\equiv{\cal D}a; \nonumber \\
X\cdot a\equiv X\cdot 
\left( 
\begin{array}{c}
w \\
a'
\end{array}
\right)=\left(
\begin{array}{c}
-w\cdot X  \\
\left[ X,a' \right]
\end{array}
\right)       
\label{covariantd}
\end{eqnarray}
with $X$ being any real skewsymmetric $n \times n$ matrix 
of one-forms. Indeed, one has the identities:     
\begin{eqnarray}
\Im m \left(\bar{a}{\cal D}a-\left(\bar{a}{\cal D}a\right)^T\right) 
=d \, \Im m \left(\bar{a}a\right), \nonumber \\ 
\Im m\left(\bar{a}\left(X\cdot a\right)-
\left(\bar{a}\left(X\cdot a\right)\right)^T \right) =
\left[X,\Im m \left(\bar{a}a\right)\right],
\label{xproperty}
\end{eqnarray}
so that both the ordinary exterior differential as well as 
the $O(n)$ connection part lead to vanishing contributions 
inserted into the imaginary part of eq. 
(\ref{newfermconstraint}) as long as $a$ satisfies the ADHM 
constraint (\ref{constraints}). For an arbitrary $O(n)$ Lie 
algebra valued one form $X$ holds 
\begin{equation}
\Re e \left(\bar{a}\left(X\cdot a\right)-
\left(\bar{a}\left(X\cdot a\right)\right)^T \right)=-{\bf L}\cdot X 
\nonumber 
\end{equation}
with ${\bf L}$ as introduced in eq. (\ref{loperator}). This allows 
to choose a connection s.t. also the remaining real part of 
(\ref{newfermconstraint}) vanishes which is found to be given by 
\begin{equation}
X={\bf L}^{-1}\cdot \Re e \left(\bar{a}da-
\left(\bar{a}da\right)^T\right).
\label{xchoice}
\end{equation}    
We assume now that, the bosonic ADHM moduli $a$ satisfy the 
constraints (\ref{constraints}), (\ref{rankcondition}) and the gauge 
fixing condition 
\begin{equation}
Y_{ik}\equiv \Re e \left(\bar{a}a\right)_{ik}= 0; \,\,\,\, \mbox{for} \, \, 
i\neq k. \nonumber 
\end{equation}
The fermionic variables are determined by eq.'s (\ref{covariantd}) and 
(\ref{xchoice}) and satisfy therefore the constraints 
(\ref{newfermconstraint}). The Jacobian factors arising from the 
integration of the bosonic $\delta$-functions in eq. (\ref{measure}) 
and the imaginary projections of the fermionic $\delta$-functions, eq. 
(\ref{fermdelta}), cancel each other as a consequence of the 
relations (\ref{xproperty}). From the real part of the fermionic 
$\delta$-functions survives the $SO(n)$ Haar measure (
multiplied by ${\bf L}$)
\[
\int_{g\in SO(n)}\prod_{i<j}\left({\bf L}\left(g^{-1}dg\right)\right)_{ij} 
=Vol(SO(n)) \det {\bf L}.
\]  
The factor $\det {\bf L}$ from the last integration drops out 
together with the $1/\det {\bf L}$ factor of the integration 
of the auxiliary variables ${\cal A}_{tot}$ in (\ref{measure}).
Viewing the induced action $S^{(n)}$ as a  mixed differential form 
\[ S^{(n)}=S^{(n)}(a,{\cal D}a) \]
we are all in all lead to rewrite eq. (\ref{modspaceint}) as 
\begin{equation}
{\cal F}_n\simeq C_n \frac{Vol(SO(n))}{Vol(O(n))}
\int_{{\cal M}_n} e^{-S^{(n)(a,{\cal D}a)}}
\label{newmodspaceint1}
\end{equation}         
with ${\cal M}_n$ denoting the $n$-instanton moduli space. The 
exponential under the integral has to be expanded s.t. the top 
form on ${\cal M}_n$ is reached. 
     
\subsection{$S^{(n)}$ as an equivariantly exact form}

The integrand on the r.h.s. of eq. (\ref{newmodspaceint1}) is also 
invariant under a $U(1)$ symmetry (besides its $O(n)$ invariance), 
the remainder of the original $SU(2)$ gauge group. We want to introduce 
an equivariant calculus with respect to this $U(1)$ symmetry, (for a 
detailed account of the equivariant differential calculus one may 
consult chapter 7 of \cite{berline}). 

Let ${\bf M}$ denote a manifold with a continuous group ${\bf G}$ 
acting on it. Vectors $X$ of the Lie algebra ${\bf g}$ of ${\bf G}$ 
are mapped to vector fields $L_X$ on ${\bf M}$. With $i_X$ we denote the 
nilpotent operation of contraction of the vector field $L_X$ with 
differential forms on  ${\bf M}$. Following Cartan \cite{cartan} 
one introduces the ``equivariant'' external differential 
\begin{equation}
d_X =d-i_X.
\label{equivdiff}
\end{equation}
$d_X$ is nilpotent on ${\bf G}$-invariant differential forms since 
one has 
\begin{equation}
d_X^2 =-\left(d \circ i_X + i_X \circ d \right)={\cal L}_X  
\end{equation}             
with ${\cal L}_X$ being identified according to Cartan's homotopy 
formula with the Lie derivative along the vector field $L_X$. The 
extension of the formalism to the setting of a vector bundle over 
${\bf M}$ is straightforward. 

Let ${\cal D}=d+A$ denote a covariant derivative acting on 
sections of such a bundle (in the case under consideration a bundle 
associated to $O(n)$). The equivariant operation ${\cal D}_X$ is 
defined in analogy to eq. (\ref{equivdiff}) by 
\begin{equation}
{\cal D}_X ={\cal D}-i_X
\label{covequivdiff}.
\end{equation} 
This gives rise to the notion of equivariant curvature 
\begin{equation}
{\cal F}_X\left(\cdot \right) =\left({\cal D}_X^2 +{\cal L}_X \right)
\left(\cdot \right)
\label{equivcurvature}
\end{equation} 
satisfying the equivariant Bianchi identity
\begin{equation}
{\cal D}_X {\cal F}_X=0
\label{equivbianchi}. 
\end{equation}
The $U(1)$ symmetry in question is represented infinitesimally 
on the bosonic ADHM parameters $w$, $a'$ by   
\begin{equation}
\delta w\sim vw, \, \, \, \, \delta \bar{w}\sim -\bar{w} v , 
\, \, \, \, \delta a'=0, 
\label{u1action} 
\end{equation}  
$v$ being the breaking parameter introduced above. Let $L_v$ denote 
the vector field on the moduli space corresponding to the 
transformation laws (\ref{u1action}). The contraction operation 
associated to this vector field is given by
\begin{equation}
i_v \cdot d w = vw, \, \, \, \, i_v \cdot d \bar{w} = -\bar{w} v , 
\, \, \, \, i_v \cdot d a'=0. 
\label{contraction} 
\end{equation}    
A simple calculation reveals that the induced action $S^{(n)}$ can 
be represented as equivariant external exterior derivative of 
an $U(1)$ invariant one form denoted by $\omega$ 
\begin{eqnarray}
\label{eq:newaction} 
S^{(n)}={\cal D}_v \cdot \omega \equiv d_v \omega \\
{\cal D}_v =d+X-i_v, \nonumber \\
d_v =d-i_v , \nonumber \\
\label{eq:omega}
\omega = -\sum_{i=1}^n 2\Re e\left( \bar{w}_i \bar{v}{\cal D} w_i\right)
\end{eqnarray}  
The two alternative representations of $S^{(n)}$ \footnote{
$S^{(n)}$ in eq. (\ref{eq:newaction}) differs by an irrelevant 
overall normalization factor from the DKM action, eq. (\ref{action}).
} 
in 
(\ref{eq:newaction}) are a consequence of the fact that the 
one-form $\omega$ is $O(n)$ invariant.

So, ${\cal F}_n$ can be represented as the integral of an 
exponential of an equivariantly exact form (superseding eq. 
(\ref{newmodspaceint1}) ) 
\begin{equation}
{\cal F}_n
\simeq \int_{{\cal M}_n} e^{-d_v\omega}.
\label{newmodspaceint2}
\end{equation}  
To demonstrate the naturalness of the equivariant calculus in 
the present context (apart from the concise appearance of the 
action (\ref{eq:newaction}), (\ref{eq:omega})) we quote the 
antichiral supersymmetry transformations induced from ${\bf N}=2$ 
field theory  to the ADHM supermoduli (as shown in \cite{dorey2})
\footnote{we quote here a special supersymmetry transformation. 
The general transformations are of the form $Q^i_{\dot{\alpha}}
a_{\alpha \dot{\beta}} =\epsilon_{\dot{\alpha}\dot{\beta}}
\left({\cal D}a\right)^i_{\alpha}
, \, \, \, i=1,2$ etc.}: 
\begin{eqnarray}
\delta a_{\alpha \dot{\beta}}=\left({\cal D}_v a\right)_{\alpha \dot{\beta}} 
\, \, \, \, \left(\equiv {\cal D} a_{\alpha \dot{\beta}}\right); 
\nonumber \\
\delta \left({\cal D}_v a\right)_{\alpha \dot{\beta}}= 
{\cal D}_v^2 a_{\alpha \dot{\beta}}= {\cal F}_v a_{\alpha \dot{\beta}} 
\equiv 
\left(
\begin{array}{c}
-\omega_{\alpha \dot{\beta}} {\cal F}_v \\
\left[{\cal F}_v,a'_{\alpha \dot{\beta}} \right] 
\end{array}
\right).
\end{eqnarray}
The closure of this representation of supersymmetry transformations - 
modulo $O(n)$ transformations - is a consequence of the equivariant 
Bianchi identity, eq (\ref{equivbianchi}).  
                                
\section{${\cal F}_n$ as a formal intersection number}\label{sec:IV}

Standard localization theory for exact equivariant forms 
\cite{berline}, \cite{schwarz} might suggest that the integral 
(\ref{newmodspaceint2}) can be localized at the set of critical 
points of the vector field $L_v$, that is , at $w_1=\cdots w_n=0$. 
This turns out not to be the case, as the residuum in question at 
the locus $w=0$ vanishes. It should also be noted that the standard 
theory, tailored for compact manifolds without boundaries, does 
not obviously apply to our problem since there are at least three 
potential obstacles \\
(i) The variables $w_i$ and $a'_{ij}$ reach out to infinity; \\
(ii) We cannot ignore a Gribov horizon, as noted above, which 
supplies one type of boundary. \\
(iii) The other type of boundary, the Donaldson - Uhlenbeck boundary 
\cite{donaldson}, \cite{uhlenbeck}, appears at places where the 
rank condition (\ref{rankcondition}) is violated. \\
To deal with item (i) we introduce a scaling variable by setting 
$a=R\hat{a}$, i.e.  
\begin{equation}
w=R\hat{w}, \, \, \, \, \, a'_{ij}=R\hat{a}'_{ij}
\end{equation}
s.t. holds \footnote{The appearance of the quadratic form on 
the l.h.s. of eq. (\ref{scalefixing}) is irrelevant as long as 
it is non-degenerate.}
\begin{equation}
\sum_{i=1} \left| \hat{w}_i \right|^2 +\sum_{i,j=1}^n 
\left| \hat{a}'_{ij} \right|^2 = 1
\label{scalefixing} 
\end{equation}       
$S^{(n)}$ reads in terms of the variables $R$, $\hat{a}$ as 
\begin{equation}
S^{(n)}=dR \, \hat{\omega}^{(n)}+Rd_v \hat{\omega}^{(n)}
\end{equation}
with the notation $\hat{\omega} \equiv \omega \left(\hat{a}, 
d \hat{a} \right)$. The scaling variable $R$ can be 
integrated straightforwardly, 
\begin{eqnarray}
{\cal F}_n
\simeq \int_{{\cal M}_n} e^{-dR \hat{\omega}-
Rd_v \hat{\omega}}=-\int_{\hat{{\cal M}}_n}\int_0^{\infty} dR 
\hat{\omega}e^{-Rd_v \hat{\omega}}=\int_{\hat{{\cal M}}_n} 
\frac{\hat{\omega}}{d_v \hat{\omega}}.  
\label{newmodspaceint3}
\end{eqnarray}
($\hat{{\cal M}}_n$ is the manifold of the rescaled moduli $\hat{a}$ )

Using the notation $\rho =\hat{\omega}/i_v \cdot \hat{\omega}$ 
we rewrite the previous equation in the form 
\begin{equation}
{\cal F}_n \simeq \int_{\hat{{\cal M}}_n} \rho \left(d\rho \right)^{4n-3}.
\label{restmodspaceint1}
\end{equation}  
To verify the equality of (\ref{newmodspaceint3}) and 
(\ref{restmodspaceint1}) one has to expand the denominator in the 
integrand of (\ref{newmodspaceint3}) and to take into account the 
identities 
\begin{equation}
\frac{\hat{\omega}}{i_v \cdot \hat{\omega}} 
\left(\frac{d\hat{\omega}}{i_v \cdot \hat{\omega}} \right)^k =
\frac{\hat{\omega}}{i_v \cdot \hat{\omega}} 
\left(d\, \frac{\hat{\omega}}{i_v \cdot \hat{\omega}} \right)^k \equiv 
\rho \left(d\rho \right)^k 
\end{equation} 
for any nonnegative integer $k$. 
The most important properties of the forms $\rho$, $d\rho$ are contained 
in the equations 
\begin{eqnarray}
\label{eq:rhocontraction}
i_v \cdot \rho =1; \\
\label{eq:drhocontraction}
i_v \cdot d\rho =0,
\end{eqnarray}
the latter equality being a consequence of the relations 
\begin{equation}
\rho=\frac{\hat{\omega}}{i_v \cdot \hat{\omega}}, \,\, 
{\cal L}_v \hat{\omega}\equiv \left(i_v \circ d + 
d \circ i_v \right) \hat{\omega}=0.
\end{equation}
One may view $\hat{{\cal M}}_n$ as a $S^1$ bundle vis-\`{a}-vis 
the action of the $U(1)$ symmetry. Eq. (\ref{eq:drhocontraction}) means 
that $\rho$ contains with coefficient $1/2|v|$ the differential of an 
angle parameterizing the $U(1)$ group orbits. $2|v|\rho$ is, with other words, 
an angular one-form of the $S^1$-bundle and $2|v|d\rho$ 
would have to be identified with the Euler class of that bundle were there 
no boundaries. The differential of the $U(1)$ angle, call it $\varphi$, 
only shows up in $\rho$ but not, according to eq. 
(\ref{eq:drhocontraction}), in $d\rho$. We may therefore substitute
$\rho$ in the integrand on the r.h.s. of eq. 
(\ref{restmodspaceint1}) by $2|v|d\varphi$ (other parts of $\rho$ lead to
vanishing contributions). Integrating out $\varphi$ we arrive at 
\begin{equation}
{\cal F}_n \simeq \int_{\hat{{\cal M}}'_n} \left(d\rho \right)^{4n-3}, 
\label{restmodspaceint2}
\end{equation}       
where $\hat{{\cal M}}'_n$ denotes the base 
of the $S^1$-bundle of which $\hat{{\cal M}}_n$ is the total bundle 
space. ${\cal F}_n$ as represented in eq. (\ref{restmodspaceint2}), 
appears as an intersection number, i.e. the integral of the 
$(4n-3)$-fold product of $d\rho$.

{\bf Remarks:} \\
1)The concrete coordinatization of the $U(1)$ group requires the choice 
of a concrete, complex scalar $U(1)$ representation. It seems 
advisable to choose an $O(n)$ invariant combination of variables 
forming such a representation. (Only in this case the $U(1)$ angle 
will not be changed along $O(n)$ orbits and, hence, can be naturally 
defined in the space $\hat{{\cal M}}_n$.)
One may, for example, combine the quaternionic components of 
\begin{eqnarray}
w=\left(w_1, \cdots ,w_n \right) \\
w_i= \sum_{\mu=0}^3 e_{\mu} \cdot w_i^{\mu},
\end{eqnarray} 
 into complex eigenvectors $\psi_i$, $\chi_i$ of the $U(1)$ 
transformations 
\begin{eqnarray}
\psi_i=w_i^0+iw_i^3; \nonumber \\
\chi_i==w_i^2+iw_i^1 \nonumber 
\end{eqnarray}
and build with them $O(n)$- invariant variables 
\begin{eqnarray}
Z=\sum_{i=1}^n  \psi_i^2 \nonumber \\
Z'=\sum_{i=1}^n  \chi_i^2 \nonumber .
\end{eqnarray}
The real and imaginary part of $Z$ ($Z'$) are of the type of doublets 
we are looking for. Suppose we parameterize the $Z$- plane with a 
radial and an angular variable. The latter may be be identified 
with the above variable $\varphi$. One is now confronted with three 
kinds of boundaries, the Gribov horizon, the Donaldson-Uhlenbeck 
(DU) boundary and the submanifold given by $Z=0$, where the 
chosen coordinatization of the $U(1)$ symmetry breaks down.  
The Gribov horizon does not give a contribution because there at 
least one $O(2)$ subgroup of $O(n)$ is restored as symmetry and 
this is not covered by the locally $O(n)$ invariant combinations 
of differential forms showing up in $S^{(n)}$. The vanishing 
of the contributions at the DU boundary is the consequence of simple 
estimates. We conclude that the integral (\ref{restmodspaceint2}) 
is localized at $Z=0$ 
\begin{equation}
{\cal F}_n \simeq \int_{\hat{{\cal M}}'_n \cap \{Z=0\}} 
\rho \left(d\rho \right)^{4n-4}. 
\label{restmodspaceint3}
\end{equation}          
We can iterate this procedure choosing as new $U(1)$ variable, e.g., the 
argument of the complex variable $Z'$ and afterwards possibly 
other combinations of variables. Unfortunately we have not been 
able to execute the recursion efficiently enough to reach a 
determination of ${\cal F}_n$ for general $n$. \\ 
2) We want to emphasize that the representation (\ref{restmodspaceint2}) 
for ${\cal F}_n$ is by no means unique because of the quasi-cohomological 
character of the problem. To illustrate the point we show that the 
result for ${\cal F}_n$ is to a large extent independent of the choice 
of an $O(n)$- connection. \\ 
Let ${\bf L}_(\alpha, \beta)$ denote the operator 
\begin{eqnarray} 
{\bf L}_(\alpha, \beta)\left(\cdot \right) = 
\alpha \left\{W, \cdot \right\} +\beta \left(
\bar{a}'\left[a',\cdot \right]- \left[\bar{a}',\cdot \right]a'\right) 
\label{deformedl}
\end{eqnarray}
with $\alpha$ and $\beta$ being positive real numbers. 
${\bf L}_(\alpha, \beta)$ is invertible on $\hat{{\cal M}}_n$. 
We may consider the modified equivariant connection
\begin{eqnarray}
{\cal D}_v^{(\alpha, \beta)} &=&d+A^{(\alpha, \beta)}-i_v, \nonumber \\
A^{(\alpha, \beta)}&=&\frac{1}{{\bf L}_(\alpha, \beta)}\Re e \left\{
\alpha \bar{w}dw+\beta \bar{a}'da'-\left(\cdots \right)^T
\right\} \nonumber
\end{eqnarray}
and the modified action 
\begin{equation}
S^{(n)(\alpha, \beta)}=-d_v \cdot \Re e\left( \bar{w}\bar{v}
{\cal D}_v^{(\alpha, \beta)}w\right)  \nonumber
\end{equation}
and 
\begin{equation}
{\cal F}_n^{(\alpha, \beta)}\simeq \int_{{\cal M}_n}
e^{-S^{(n)(\alpha, \beta)}}.  \nonumber 
\end{equation}
We want to show that ${\cal F}_n^{(\alpha, \beta)}={\cal F}_n$. 
In fact we have 
\begin{eqnarray}
{\cal F}_n -{\cal F}_n^{(\alpha, \beta)}=\int_{{\cal M}_n} 
\left(e^{-S^{(n)}}-e^{-S^{(n)(\alpha, \beta)}}\right)= \nonumber \\
\int_{{\hat{{\cal M}}_n}} \left(\frac{ \Re e \left(\bar{w}\bar{v}
{\cal D}_v w\right)}{d_v \cdot  \Re e\left(\bar{w}\bar{v}
{\cal D}_v w\right)}-\frac{ \Re e \left(\bar{w}\bar{v}
{\cal D}_v^{(\alpha, \beta)}w\right)}{d_v \cdot \Re e 
\left(\bar{w}\bar{v}{\cal D}_v^{(\alpha, \beta)}w\right)}\right). \nonumber
\end{eqnarray}
The topform (t.f.) of the difference of terms in the last integral is exact, 
as can be inferred from the following chain of identities:  
\begin{eqnarray}
\left\{
\frac{ \Re e \left(\bar{w}\bar{v}
{\cal D}_v w\right)}{d_v \cdot  \Re e\left(\bar{w}\bar{v}
{\cal D}_v w\right)}-\frac{\Re e \left(\bar{w}\bar{v}
{\cal D}_v^{(\alpha, \beta)}w\right)}{d_v \cdot  \Re e 
\left(\bar{w}\bar{v}{\cal D}_v^{(\alpha, \beta)}w\right)}
\right\}_{t.f.}= \nonumber \\
\left. 
\frac{
\Re e \left(\bar{w}\bar{v}
{\cal D}_v w\right) d_v \Re e 
\left(\bar{w}\bar{v}{\cal D}_v^{(\alpha, \beta)}w\right)- 
d_v \Re e\left(\bar{w}\bar{v}
{\cal D}_v w\right) \Re e \left(\bar{w}\bar{v}
{\cal D}_v^{(\alpha, \beta)}w\right)}
{d_v \Re e\left(\bar{w}\bar{v}
{\cal D}_v w\right) \cdot d_v \Re e 
\left(\bar{w}\bar{v}{\cal D}_v^{(\alpha, \beta)}w\right)}  
\right|_{t.f.}= \nonumber \\
\left. d_v \cdot 
\frac{\Re e \left(\bar{w}\bar{v}
{\cal D}_v w\right) \Re e 
\left(\bar{w}\bar{v}{\cal D}_v^{(\alpha, \beta)}w\right)}
{d_v \Re e\left(\bar{w}\bar{v}
{\cal D}_v w\right) \cdot d_v \Re e 
\left(\bar{w}\bar{v}{\cal D}_v^{(\alpha, \beta)}w\right)}
\right|_{t.f.} = \nonumber \\
\left. d \cdot 
\frac{\Re e \left(\bar{w}\bar{v}
{\cal D}_v w\right) \Re e 
\left(\bar{w}\bar{v}{\cal D}_v^{(\alpha, \beta)}w\right)}
{d_v \Re e\left(\bar{w}\bar{v}
{\cal D}_v w\right) \cdot d_v \Re e 
\left(\bar{w}\bar{v}{\cal D}_v^{(\alpha, \beta)}w\right)}
\right|_{t.f.},
\label{difference}
\end{eqnarray}
where for the last identity use has been made of the fact that the 
equivariant and ordinary exterior derivatives coincide if evaluated 
on the top form.
One concludes from arguments as used above in remark (i) that there 
are no boundary contributions to the integral of the exact form 
(\ref{difference}) and hence ${\cal F}_n ={\cal F}_n^{(\alpha, \beta)}$.

\section{Punctured Riemann spheres and instantons}\label{sec:V}

Matone, \cite{matone3}, has pointed to an interesting analogy between 
the recursive determination of the Weil-Peterson volume of the 
moduli space of punctured Riemann sphere and the calculation of 
the ${\cal N}=2$ prepotential. The analogy has a ``macroscopical'' 
(in the sense used in the introduction) and a microscopical aspect. 
While the former is very neat the later is so far of a purely 
speculative nature. Let us start to describe the macroscopic approach 
towards the determination of the SW coefficients by Matone 
\cite{matone3} \footnote{
A more general one-term recursion relation, which 
is exceedingly complicated, has been found by Chang and d'Hoker 
\cite{chan}. We belive that a microscopical reconstruction 
of this recursion is more remote than that of Matone's. 
}.

Let $u$ denote the $SU(2)$ invariant order parameter of the 
${\cal N}=2$ Higgs-Yang-Mills theory:
\[
u=\langle tr \Phi^2\rangle.
\]
Superconformal Ward identities can be used to derive, \cite{howe}, 
the following 
relation between the prepotential and $u$, both being considered as 
functions of $v$:
\[
\left\{{\cal F}(v)-\frac{1}{2}v\frac{\partial}{\partial v}{\cal F}(v)
\right\}=\frac{1}{2}u(v).
\]
The variable $v$ and its dual 
\[
v_D\equiv \frac{\partial {\cal F}(v)}{\partial v}
\]
appear in the construction of SW as period integrals of a 
hyperelliptic Riemann surface:
\[
v_D=\frac{\sqrt{2}}{\pi}\int_1^u \frac{dx\sqrt{x-u}}{\sqrt{x^2-1}},
\,\, v=\frac{\sqrt{2}}{\pi}\int_{-1}^1 \frac{dx\sqrt{x-u}}{\sqrt{x^2-1}};
\]
$v$ and $v_D$ may alternatively be represented as solution system of a 
Fuchs differential equation (a hypergeometric equation in the case at 
hand)
\begin{equation}
\left\{
\left(1-u^2\right)\partial_u^2-\frac{1}{4}
\right\}v_D=0.
\label{fuchs}
\end{equation}
Inverting the functional dependence Matone \cite{matone3} derived  
a non-linear differential equation for $u$ as function of $v$,
\begin{equation}
\left(1-{\cal G}^2\right){\cal G}''+\frac{1}{4}ag'^3=0
\label{nonlinear}
\end{equation}
with the notation \[
{\cal G}=\frac{-\pi i }{2} \, u\equiv \pi i \left({\cal F}(v)-\frac{1}{2}v 
\frac{\partial}{\partial v} {\cal F}(v)
\right).
\]
Inserting the power series 
\[
{\cal G}(v)=\sum_{i=1}^{\infty} {\cal G}_n v^n
\]
into (\ref{nonlinear}) one obtains a three term recursion relation 
for the coefficients ${\cal G}_n$;
\begin{eqnarray}
{\cal G}_{n+1}=\frac{1}{2(n+1)^2}\left\{(2n-1)(4n-1){\cal G}_n \right. 
\qquad \qquad \qquad \qquad \qquad \quad \nonumber \\ 
\left. +\sum_{k=0}^{n-1}c(k,n){\cal G}_{n-k}{\cal G}_{k+1}-2\sum_{j=0}^{n-1}
\sum_{k=0}^{j+1}d(j,k,n) {\cal G}_{n-j}{\cal G}_{j+1-k}{\cal G}_k,
\right\}
\label{recursion}
\end{eqnarray}
($c(k,n)=2k(n-k-1)+n-1$, \, \, $d(j,k,n)=[2(n-j)-1][2n-3j-1+2k(j-k+1)]$) 
and therewith also a recursion relation for the coefficients 
${\cal F}_n= {\cal G}_n /2\pi in$. 

The moduli space of a Riemann sphere with $n$ punctures 
\[
\Sigma_{0,n}=\hat{C}\setminus \left\{z_1, \cdots ,z_n; z_i 
\neq z_j \,\,\, \mbox{for} \,\,\,  i\neq j
\right\},
\]
with $\hat{C}$ denoting the compactified complex plane, is the 
space of isomorphism classes of punctured spheres 
\[
{\cal M}_{0,n}=
\left\{z_1, \cdots ,z_n; z_i \neq z_j \,\,\, \mbox{for} \,\,\,  i\neq j
\right\}/\left\{Symm(n)\times PSL(2,C)\right\},
\]
Let $\omega_{WP}^{(n)}$ be the two-form  on ${\cal M}_{0,n}$ 
associated with the Weil-Petersson metric giving rise to a finite 
W-P-volume 
\begin{equation}
Vol_{WP}\left({\cal M}_{0,n}\right)=\frac{1}{(n-3)!}\int_{{\cal M}_{0,n}}
\omega_{WP}^{n-3}.
\label{wpvolume}
\end{equation} 
Zograf proves \cite{zograf} that the quantities 
\[
v_n =\frac{(n-3)!}{\pi^{2(n-3)}} \,Vol_{WP}\left({\cal M}_{0,n}\right)
\]
obey for $n\geq 4$ the recursion relation 
\begin{eqnarray}
v_n=\sum_{j=1}^{n-3}\frac{j(n-j-2)}{n-1}
\left(
\begin{array}{c}
n-4 \\
j-1
\end{array}
\right)
\left(
\begin{array}{c}
n \\
j+1
\end{array}
\right)
v_{j+2}v_{n-j}; \, \, \, \, v_3=1.
\label{zografrecurence}
\end{eqnarray} 
The origin of the two-term structure of (\ref{zografrecurence}) 
is easy to understand within Zograf's approach to the problem. He 
is executing the integral of one of the $n-3$ two forms in 
(\ref{wpvolume}). The homology cycles separates the punctures 
on $\hat{C}$ into two groups leading 
to a factorization into the product of the volumes of two spheres, 
each with a smaller number of punctures, as a result of the 
localization of the two-dimensional integration on the Deligne, 
Knudsen, Mumford boundary \cite{deligne} of ${\cal M}_{0,n}$. 
This concludes our rough sketch of the microscopic part of the 
problem.

To find the analogues of eq.'s (\ref{fuchs}), (\ref{nonlinear}) we 
introduce, following Matone \cite{matone2}, the generating function 
\begin{eqnarray}
g(x)=\sum a_k x^{k-1}; \nonumber \\
a_k=\frac{v_k}{k-1}\left((k-3)!\right)^2. \nonumber
\end{eqnarray}
The recursion relation for the coefficient $g_n$ is equivalent to 
the non-linear differential equation for the function $g$, \cite{matone2} 
\begin{equation}
x(x-g)g''=xg'^2+(x-g)g'.
\label{nonlinear1}
\end{equation}
With an appropriate change of variables one achieves a functional 
inversion of (\ref{nonlinear1}), \cite{kontsevich}, \cite{kmz} with 
the linear substitute for 
(\ref{nonlinear1}), a Bessel differential equation,
\begin{equation}
y\frac{d^2x}{dy^2}+x=0.
\label{linear1}
\end{equation}
The analogies between (\ref{wpvolume}) and (\ref{restmodspaceint2}), 
(\ref{linear1}) and (\ref{fuchs}), (\ref{nonlinear1}) and 
(\ref{nonlinear}) are obvious and striking. What is missing for a 
perfect analogy is a transposition of the above quoted work of 
Zograf to the instanton problem. It is tempting to speculate that 
that the integral for ${\cal F}_n$ eq. (\ref{restmodspaceint2}), 
becomes localized on (some part of) the Donaldson-Uhlenbeck boundary. 
But this is at least for the representation given by 
eq.(\ref{restmodspaceint1}) not the case. The residua in question do 
vanish. 

\section{Discussion}

Keeping the above impressive analogies in mind we have nevertheless 
to emphasize that the topological complexity of the instanton 
manifold is by no means to be compared with the clear situation 
encountered in the case of punctured Riemann spheres. The 
ADHM moduli manifold may be considered as a non-trivial 
principal $O(n)$- bundle  
(apart from being  a principal $U(1)$-bundle which was of prime 
importance above). The 
non-triviality is reflected in the fact that an $O(n)$ gauge 
fixing condition has to be chosen together with a domain of  
integration bounded by a Gribov horizon. Configurations of clusters 
of ``instantons of small size'' on the Donaldson-Uhlenbeck boundary 
of which one might speculate that they give rise to the recursion 
mentioned in the preceeding section, are orbifold points of the 
$O(n)$ principal bundle. We have not been able to verify the 
latter speculation on the basis of the representation given 
by eq. (\ref{restmodspaceint2}) since the relevant residua vanish. 
A crucial 
point for any further progress in this matter will be - to our 
opinion - that one finds a (so far missing) idea as to which geometrical 
fact the three-term recursion could be attached. (In the case 
of punctured spheres it is the effectiveness of a lasso as 
a tool for catching whatever one wants to catch in two dimensions.)
For that purpose a detailed knowledge of second cohomology classes 
$H^2$ of the space $\hat{{\cal M}}'_n$ would be of prime importance. 
The complete analogy between punctured Riemann spheres and 
the SW prepotential in four dimensions, enthusiastically announced 
in \cite{matone2} has still to be found. We hope that our 
representation (\ref{restmodspaceint2}) of the coefficients 
${\cal F}_n$ as formal 
intersection numbers may be a useful step into this direction.

\section*{Acknowledgments}
R.F. owes his initiation to the subject of the Seiberg-Witten 
potentials to the late Lochlainn O'Raifeartaigh and to Ivo Sachs. 
This is gratefully  acknowledged.
R.P. acknowledges the partial financial support of INTAS 00-561 
and of the Alexander von Humboldt foundation of Germany.

\vfill\eject

\end{document}